\documentclass[12pt]{article}
\usepackage{amsmath}
\usepackage{epsfig}
\usepackage{amstext,amssymb,amsfonts}

\date{January 2011}

\advance\voffset by -2.0cm
\advance\hoffset by -1.75cm
\def\be{\begin{equation}}
\def\ee{\end{equation}}
\def\ba{\begin{eqnarray}}
\def\ea{\end{eqnarray}}

\def\Tr{{\rm Tr}}

\def\IZ{{\mathbb{Z}}}

\def\CC {{\cal C}}

\def\CN {{\cal N}}

\def\CF {{\cal F}}

\def\CZ {{\cal Z}}

\def\CZ{{\cal Z}}

\renewcommand{\Im}{{\rm Im }}
\renewcommand{\Re}{{\rm Re }}

\newcommand{\eg}{{\it e.g.~}}
\newcommand{\ie}{{\it i.e.~}}

\title{\vspace{-1cm}\begin{flushright}{\small CALT-68-2818}\end{flushright}\vspace{2cm}
\LARGE Phase transitions in symmetric orbifold CFTs and universality
}

\author{
Christoph A.~Keller\thanks{\tt E-mail: ckeller@theory.caltech.edu} 
\\ \\
{
California Institute of Technology,}\\
\vspace*{-0.1cm} {
Pasadena, CA 91125, USA }
}

\begin{document}

\maketitle

\begin{center}
{\bf Abstract}
\end{center}
Since many thermodynamic properties of black holes
are universal, the thermodynamics of their 
holographic duals should be universal too. 
We show how this universality
is exhibited in the example of 
symmetric orbifolds of general two dimensional CFTs. 
We discuss the free energies and phase diagrams of 
such theories and show that they 
are indeed universal in the large $N$ limit.
We also comment
on the implications of our results 
for the classification of CFTs that
can have an interpretation as holographic
duals to gravity theories on $AdS_3$.

\newpage
\renewcommand{\theequation}{\arabic{section}.\arabic{equation}}


\section{Introduction and overview}
\subsection{Introduction}
Investigating the thermodynamic properties of black holes
has a long history. A common theme in these
investigations is that many black holes are characterized
by very few parameters, and that their thermodynamic 
properties are universal.
Another theme is the search for an underlying microscopic
description of the thermodynamics. A very fruitful
approach to this has been the application
of the AdS/CFT correspondence \cite{Maldacena:1997re}.
In view of this, it should be possible to see
this universality also from the holographic dual
theory of the black hole.

In this note we will concentrate on one specific class
of such holographic dual CFTs, namely symmetric orbifolds
of two dimensional CFTs.
In many string configurations leading to $AdS_3$ geometries, such
as D1-D5 systems, the holographic dual conformal field theories
are given by (deformations of) symmetric orbifold CFTs \cite{Aharony:1999ti}. 
In this particular case, the D1 branes are considered as
instantons of the low energy theory on the D5 branes, 
whose moduli space is then a deformation of the
symmetric product of the moduli space of a single instanton
\cite{Vafa:1995bm,Witten:1997yu}.
It is thus
very natural to investigate thermodynamic properties of symmetric
orbifold CFTs. 

The fact that we are dealing with CFTs
in two dimensions will be very helpful for our
analysis, as it allows us to use methods
specific to $d=2$. One crucial property that
will be central to our analysis is modular invariance
of the partition function. This allows us to relate
the free energy for a different ranges of temperature
and other parameters to each other. In particular
it is then sufficient
to know the behavior of the theory at low temperature
to also obtain the high temperature behavior.

In the remainder of this section, we will make
some general remarks about the free energy
and phase transitions of 2d CFTs. In section~2
we use the methods described above to
analyze properties of the free energy
of symmetric orbifold CFTs. 
We will discuss the phase diagram
of their Hawking-Page transitions \cite{Hawking:1982dh},
the behavior near the phase transitions, and
the critical exponents for a class
of order parameters. In particular we will
show that many of those properties are indeed universal,
and do not depend on the underlying theory.

Even though our analysis is for general, not necessarily 
holomorphic theories, we will also briefly discuss the
behavior of the elliptic genus for $N=2$ theories.
In section~3 we will also discuss 
what criteria general theories that are not symmetric
orbifolds have to satisfy if they are to exhibit the same
universal behavior.

\subsection{Some general remarks on two dimensional CFTs}
Consider a conformal field theory in two dimensions with  left- and right-moving 
central charges  such that $c_L-c_R = 0\ \textrm{mod}\ 24$.  
We will 
 assume that the CFT has a discrete spectrum with finite multiplicities, and that its 
 vacuum is unique.
 Also we will only consider unitary theories, so that all states have positive
 norm and non-negative weight. 
We also define the theory 
on a circle whose radius  sets the unit  of length.
By the usual arguments of QFT at finite temperature, we can
then obtain the partition function as the vacuum amplitude
on the torus, whose modulus $\tau$ is fixed by
the temperature and the spin potential.
 
More precisely, we define the partition function as
 \be\label{gform}
Z(\tau,\bar \tau) =   {\rm Tr}  \, q^{L_0-c_L/24} \bar q^{\, \bar L_0-c_R/24}\, =\,
\sum_{m,\bar m \in I} d(m,\bar m) q^m \bar q^{\bar m}\ .
\ee
As
usual we have defined   
\be  
q = e^{2\pi i\tau} = e^{ - \beta +  i \mu}  \ . 
\ee
Here $\beta$ is the inverse temperature and the spin potential
$\mu$ is the variable conjugate to the momentum on the circle. 
We will often set $\mu = 0$ and focus on  the behavior 
of $Z$ as a function of $\beta$. 

For reasons that will become clear in a moment,
we have defined the vacuum to have energy $-c/24$.
When discussing physical quantities, it will often be useful 
to shift the vacuum energy to zero. We will denote quantities which are
shifted in such a way by a tilde. In particular
we define
\be
\tilde Z(\tau,\bar \tau) =   q^{c_L/24} \bar q^{c_R/24} Z(\tau, \bar\tau)=
\sum_{m,\bar m \in \tilde I} \tilde d(m,\bar m) q^m \bar q^{\bar m}\ ,
\ee
where  $ \tilde d(m,\bar m)$
is the number of states at left- and right-moving levels $(m,\bar m)$.  
Finally, let us define the free 
energy $F$ as
\be
Z(\tau, \bar \tau) = e^{-2\pi \Im(\tau)F(\tau,\bar \tau)} = e^{-\beta F(\tau,\bar \tau)}\ .
\ee
Similarly, define $\tilde F$ as the logarithm of $\tilde Z$.

Many  of the properties of  $Z$  follow from general principles.
As mentioned above, the partition function $Z$ corresponds to the
vacuum amplitude of the Euclidean theory on the torus.
As the theory is conformal, it only depends on the conformal
structure. The partition function $Z$ must therefore be 
invariant under the modular group $\Gamma=SL(2,\IZ)$.
Note that this holds only for $Z$, and not for the shifted
$\tilde Z$.
It follows that we only need to know $F$ in one fundamental region of $\Gamma$. For all
other values of $\tau$ we can simply apply the appropriate modular 
transformation. For example for $\mu=0$ we can use the modular transformation
$S$ to get
\be\label{modular}
 Z (\beta) =   Z (4\pi^2/\beta)\ . 
\ee
The behavior of $F$ for very low temperatures  
is dominated by the vacuum contribution, and by 
using (\ref{modular}) we can also obtain the 
behavior at very high temperature,
\be\label{asym}
  F(\beta) \simeq  \left \{ \begin{array}{c} 
  -\frac{(c_L+c_R)}{24} \,  \ \ \ \  {\rm for} \ \beta \gg 1\ ,   \\  \, \\
- \frac{c_L+c_R}{24} \, \frac{4\pi^2}{\beta^2}   \ \ \ \  {\rm for} \ \beta \ll  1 \ .    \end{array} \right.  
\ee
The behavior in the high and low temperature limit is thus universal 
and depends only on the central charges. The behavior
between this limits on the other hand
is of course not universal and
depends on the spectrum of the CFT. The only thing we
know is that it is smooth and monotonically decreasing.
One of the main results of
our analysis will be that for symmetric orbifolds
the temperature region where the behavior is universal
extends towards the self-dual temperature $\beta=2\pi$
as $N$ grows large. In the limit of $N\rightarrow \infty$,
the free energy becomes universal for all temperatures.

In fact, the modular transformation we have discussed 
can be used
to obtain the asymptotic growth
of the number of states.
For  $m$ and  $\bar m$ very large it is given by Cardy's formula, 
\be 
d(m,\bar m) \sim  \exp\left(2\pi \sqrt{\frac{c_L m}{6}}\right)
\exp\left(2\pi\sqrt{\frac{c_R \bar m}{6}}\right)\   ,  
\ee
where we have only given the leading exponential growth. 
It follows from this formula  that the
 infinite sum in (\ref{gform})  converges absolutely,  
 so that ${\rm log}  Z $ and all its derivatives  are continuous functions
of temperature.

\subsection{Hagedorn transitions and Hawking-Page transitions}
   That ${\rm log}  Z $ and all its derivatives  are continuous functions of $\beta$  is to be expected
 on even more general grounds.  A quantum field theory with finitely many local degrees of freedom
 cannot have a phase transition at finite volume. Phase transitions can, however,  occur  if the
 number of degrees of freedom is taken to infinity. In particular
 the free energy can diverge above a certain temperature,
 so that a Hagedorn transition occurs.

  We will therefore 
 look for phase transitions in  families  $\{\CC^{(N)}\}_{N\in\mathbb{N}}$ of CFTs
 whose central charges are given by $c_{L,R}(N)= c_{L,R} N$.  
Plugging this into
(\ref{modular}), we see immediately that for
large enough temperatures the free energy is
proportional to $N$, which means that for
$N=\infty$ the free free energy indeed starts to diverge
above some Hagedorn temperature $T_H$.

Let us briefly discuss how to
interpret this Hagedorn transition 
in the context of the AdS/CFT correspondence.
The natural units
on $AdS_3$ are given in terms of the AdS radius $R$. Since $R \sim N$, we 
introduce the rescaled free energy
\be
f_N(\tau) = \frac{1}{N}F_N(\tau)\ , \qquad f(\tau) = 
\lim_{N\rightarrow\infty} f_N(\tau) \ .
\ee
Note that we have not shifted the vacuum energy, so that the
energy of empty AdS space is $-(c_L+c_R)/24$. As usual,
$f_N$ is analytic, but due to the limit $f$ can
exhibit phase transitions.
Such a transition corresponds to a Hawking-Page transition: 
below the critical temperature, the free energy is dominated
by the contribution of the $AdS$-vacuum. Above it, the entropy 
of the BTZ black holes gives the dominating contribution.

\begin{figure}
\begin{center}
\includegraphics[width=0.4\textwidth]{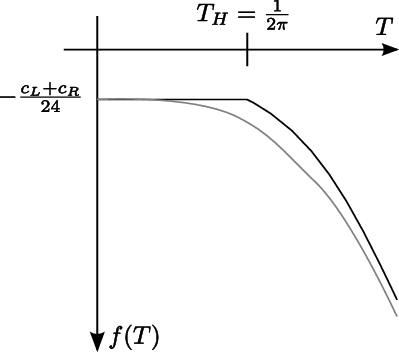}
\end{center}
\caption{\label{fig:freeEnergy}
A generic CFT (gray line) has no
phase transition and is universal only for
very high and very low temperatures. The symmetric
orbifold for $N=\infty$ (black line) is universal at all
temperatures and has a first order phase transition
at $T_H=\frac{1}{2\pi}$.}
\end{figure}

We will show that for symmetric orbifold theories 
the Hagedorn transition occurs at $T_H=\frac{1}{2\pi}$.  
In the corresponding gravity theory there is
then a Hawking-Page transition at that temperature.
More precisely, we will show that the free energy is given by
\be \label{FAdS}
f(\beta) = \left \{ \begin{array}{ccc} -\frac{c_L+c_R}{24}   &:& \beta > 2\pi \\
-\frac{4\pi^2}{\beta^2}\frac{c_L+c_R}{24}  &:& \beta < 2\pi \end{array} \right.\ .
\ee
As a general remark note that $\tilde F$ is the quantity
that appears naturally when analyzing phase transitions
of the conformal field theory in its own right, whereas
$f$ is the natural quantity for the analysis
of the CFT as a holographic dual to the gravity side.
Of course all important properties of $f$ can be 
reconstructed from $\tilde F$ and vice versa, but 
depending on the context one or the other will be
more appropriate when doing computation.

\subsection{The phase diagram}
\begin{figure}
\begin{center}
\includegraphics[width=0.4\textwidth]{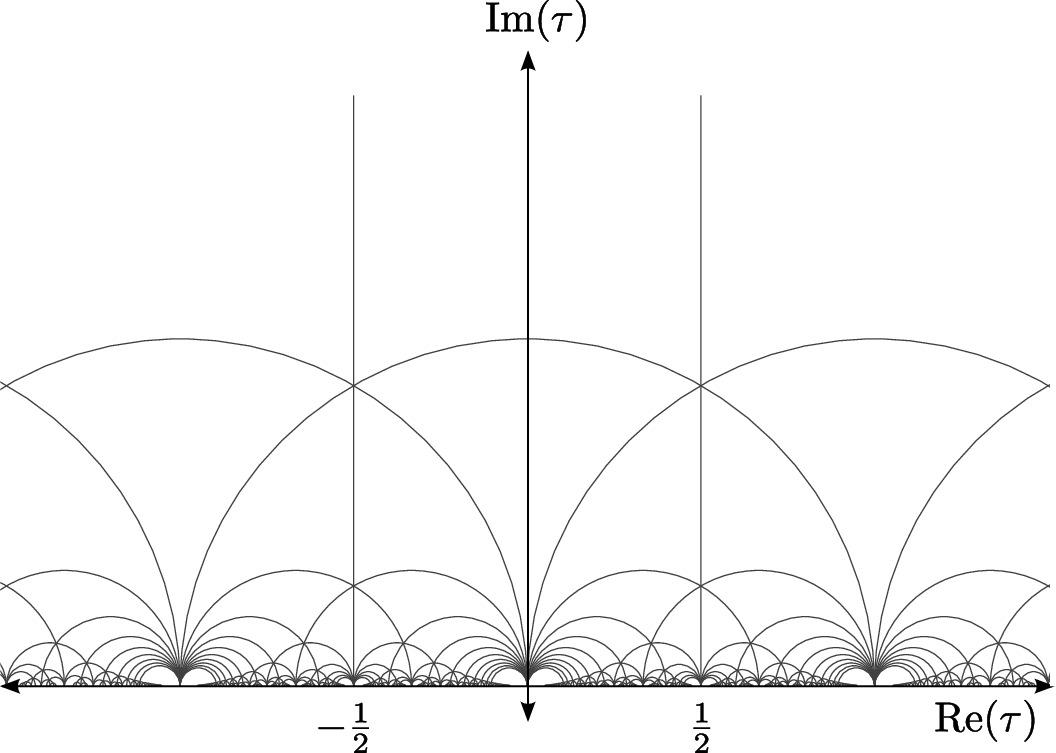}
\includegraphics[width=0.4\textwidth]{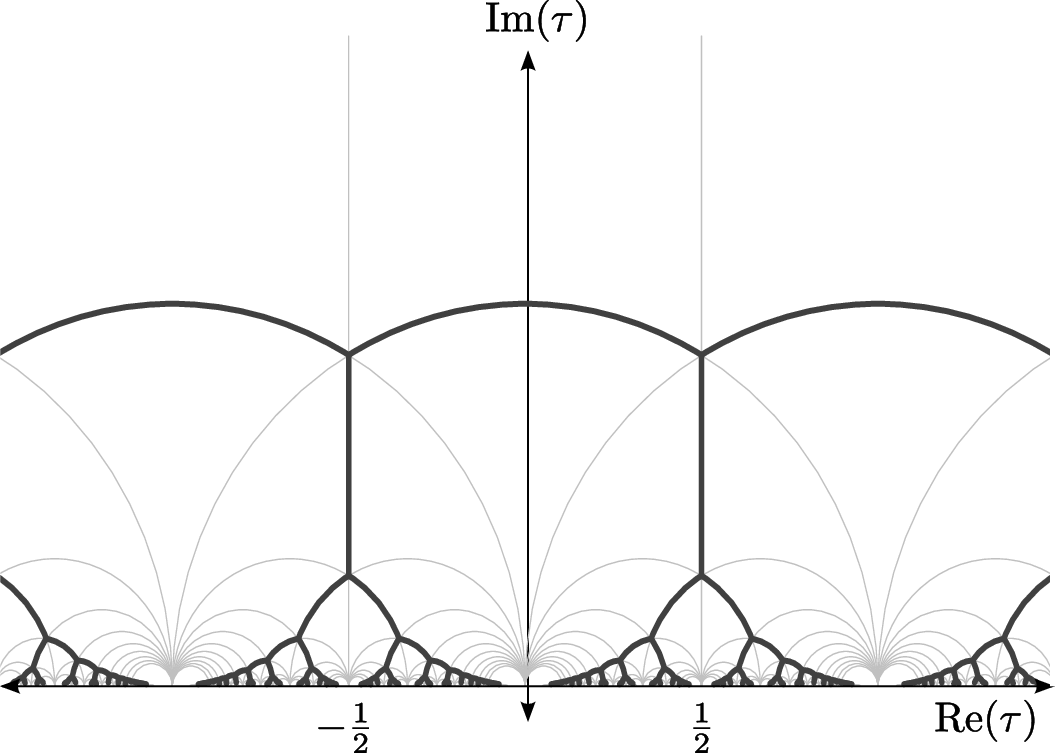}
\end{center}
\caption{\label{fundregion} 
To the left, the tessellation of the upper half plane
into fundamental regions of $SL(2,\IZ)$. To the right,
the proposed phase diagram, taking into account the
invariance under $T$. (Figures taken from \cite{Maloney:2007ud}.)}
\end{figure}

Let us briefly sketch why one expects exactly such a
Hawking-Page phase transition for 
large $N$ theories. As a toy model we take a holomorphic theory with
$c_L=24N$
and consider only the contribution $q^{-N}$ of the vacuum,
neglecting all higher weight states. 
This is clearly not a modular invariant partition function,
so to render it
modular invariant, we sum over all images of $\Gamma= SL(2,\IZ)/B(\IZ)$,
where $B(\IZ)$ is the translation subgroup generated by $T$.
We obtain \footnote{This sum is divergent and needs to be regularized. We will
ignore this issue since it does not affect the point we are making.}
\be \label{Poincare}
Z(\tau) = \sum_{(c,d)=1} e^{-2\pi i N\frac{a\tau+b}{c\tau +d}}\ . 
\ee
If we tessellate the upper half plane into fundamental regions
of $\Gamma$ as in figure~\ref{fundregion},
then
for every fundamental region there is an element 
$\left(\begin{array}{cc} a&b\\c&d\end{array}\right)\in \Gamma$ so that
the corresponding term in the sum is dominant. 
When we move from 
one fundamental region to another, a different term becomes
dominant. In particular when we send $N\rightarrow\infty$,
a phase transition will occur on the boundary
between the regions. One might thus expect the phase diagram
to look like the left 
diagram in figure~\ref{fundregion}. On the other hand
it is straightforward to see that $F$ is invariant
under $\tau \mapsto \tau+1$, so that
the phase space diagram should be given by the
right diagram in figure~\ref{fundregion}
\cite{Dijkgraaf:2000fq}. 

The Poincar\'e sum (\ref{Poincare}) has a very natural interpretation
on the gravity side. The classical solutions of Euclidean gravity with
fixed conformal boundary $\tau$ are
essentially given by BTZ black holes which correspond to
quotients of the AdS vacuum by elements
of $SL(2,\IZ)$ \cite{Maldacena:1998bw}. When evaluating
the Euclidean gravity path integral, we are instructed
to sum over all such classical solutions,
which gives the Poincar\'e sum (\ref{Poincare}). (See 
\cite{Dijkgraaf:2000fq, Manschot:2007ha} for
a more detailed discussion.) 
The structure of the phase diagram can thus be explained
by the different BTZ black holes which dominate the free energy
in different regions of phase space.

For instance, if we restrict to $\mu=0$
we can restrict the sum over images to 
 \be 
 Z_N = e^{N
  \beta} + e^{N 4\pi^2/\beta} \ \ \ .  
  \ee 
The free energy then has exactly the behavior (\ref{FAdS}).
    
There are two problems with this argument. First, it only works 
for holomorphic partition functions. We have made use of
the mathematical fact that a holomorphic modular form
can be reconstructed by taking the Poincar\'e sum (\ref{Poincare})
of its polar part. No analogue theorem exists for
non-holomorphic modular forms.

Second, a CFT necessarily contains many more states than just
the vacuum. In a family of CFTs it is thus possible that
the contribution of these states smoothes out the
phase transition coming from the vacuum term. A simple
example that shows that this can happen is the
family of $N$ tensor products of a CFT. In this
case the partition function is simply the 
$N$th power of the original partition function,
and 
\be
f(\beta)=f_N(\beta)=f_1(\beta)\ ,
\ee
which is analytic in $\beta$.

In fact, if we want the full theory to exhibit
the same behavior as (\ref{Poincare}), it is
straightforward to see that the number of
states $\tilde d_N(m,\bar m)$ in the family
has to satisfy the criterion
\be \label{criterion}
\lim_{N\rightarrow\infty} \frac{1}{N}\log \left( 1+ \sum_{m,\bar m} 
\tilde d_N(m,\bar m) e^{-2\pi(m+\bar m)}\right) = 0\ .
\ee
This follows from the requirement that the 
contributions of the higher level terms
must be subleading in $N$.
This is where the symmetric orbifold becomes important.
Starting from the tensor theory described above,
it projects out most of the low lying states,
such that the resulting theory does indeed exhibit 
phase transitions.


\section{Universal properties of large $N$ symmetric orbifolds}
\setcounter{equation}{0}
\subsection{Symmetric orbifolds} \label{ss:symOrb}
We start out with the partition function of an arbitrary seed theory $\CC_1$
(which satisfies the technical conditions outlined in the introduction),
\be
Z(\CC_1)= \sum_{m,\bar m \in I} d(m,\bar m) q^m \bar q^{\bar m} \ .
\ee
We assume that there is indeed a discrete set of states $I$,
and that lowest lying state, the vacuum, occurs exactly once. 

As we have seen above, for the family of $N$ fold tensor
products of the seed theory $\CC_1$, $f$ does not 
exhibit any phase transitions, since the higher
lying states smoothen out the free energy.
To eliminate as many of those states as possible,
we need to perform an orbifold.
As the tensor product is symmetric under permutations
of its factors, we can orbifold with respect to
the permutation group $S_N$.
This orbifold projects out most of the low lying states
of the theory.
As in any orbifold however, it also introduces new
twisted sectors that are given by the conjugacy classes
of the orbifold group. For $S_N$ these are simply
the product of cycles
\be
(1)^{N_1} (2)^{N_2} \cdots (s)^{N_s}\ , \qquad \sum_n nN_n =N\ .
\ee
In those twisted sectors we again need to project out states 
which are not invariant
under the stabilizer subgroup of the conjugacy classes,
which in this case is given by the semidirect product 
of cyclic permutations $\mathbb{Z}_n$ within a given
cycle and permutations of different cycles of the
same length. 
As is often the case in such computations, it turns
to be easier to compute the grand canonical
ensemble, \ie the generating function of
the partition functions
\cite{Dijkgraaf:1996xw,Dijkgraaf:1998zd,Bantay:2000eq}
\be \label{genF}
\mathcal{Z} = \sum_{N\geq 0} p^N Z(S^N \CC_1) = 
\prod_{n>0}\prod_{m,\bar m\in I} (1-p^n q^{m/n} \bar q^{\bar m/n})^{-d(m,\bar m)\delta^{(n)}_{m-\bar m}}
\ .
\ee
Here the periodic Kronecker symbol $\delta^{(n)}_{m-\bar m}$ is
defined to be 1 when $m-\bar m$ is a multiple of $n$, and
0 otherwise. It ensures that only local fields are kept in the
spectrum.
Alternatively, to make the modular properties of the partition functions
manifest, we can also express $\CZ$ in terms of Hecke operators,
\be\label{symHecke}
\CZ = \exp \left( \sum_{L>0} \frac{p^L}{L} T_L' Z(q,\bar q) \right) \ ,
\ee
where the action of the Hecke operator $T_L'$ is defined as
\be \label{Hecke}
T_L' f(\tau,\bar \tau) = \sum_{d|L}\sum_{b=0}^{d-1} 
f\left(\frac{L\tau+bd}{d^2},\frac{L\bar\tau+bd}{d^2} \right)\ .
\ee
We have put a prime on $T_L$ as a reminder that this definition differs
in normalization from the standard Hecke operator.

\subsection{Universal properties{\protect\footnote{The original version of this
section contained an incorrect expression for 
$\tilde d_\infty(m,\bar m)$. This version 
has been updated to reflect the correct
expressions obtained more recently in \cite{Hartman:2014oaa}.
}}}
We will now show that in the limit $N\rightarrow\infty$, these
families of theories have universal properties.
Universal means that they only depend on the central charge
of the seed theory $\CC_1$, but not on any other 
properties of its spectrum. 
We will find the following universal properties:
\begin{enumerate}
\item For $c_L=c_R$, the number of states always grows as 
$\tilde d_\infty(m,\bar m)  \sim e^{4\pi \sqrt{m\bar m}}$.
\item The divergent part of the free energy
near the Hagedorn transition is universal. 
In particular, as we will show later on, this 
implies that the critical exponent for 
the phase transitions of certain
families of order parameters is always 1.
\item The phase diagram for the rescaled free
energy $f(\tau)$ is universal, and is
given by figure~\ref{fundregion}.
\end{enumerate}
Let us now show these properties. For our analysis
it will be easiest to work with the 
shifted free energy $\tilde F$, and then translate
our results into statements about $f$ if needed.

\subsubsection*{1. Growth of states}
Let us find an expression for the growth of the number of states
of the symmetric orbifold theory of large $N$.
First, we show that the partition function $\tilde Z_\infty$
of the $N=\infty$ theory exists, \ie
the number of states $
\tilde d_\infty(m,\bar m) = \lim_{N\rightarrow\infty} \tilde d_N(m,\bar m)$
remains finite.
More precisely, we will show that if $c_L=c_R$ 
\be \label{dinfty}
\tilde d_\infty(m,\bar m)  \sim e^{4\pi \sqrt{m\bar m}}\ \qquad
{\rm for}\ m\ ,\bar m \gg c_R = c_L
\ .
\ee
This behavior is certainly consistent with a Hagedorn
transition at $\beta = 2\pi$. In fact, (\ref{dinfty}) is also
true for finite $N$, as long as $m, \bar m$ are sufficiently
smaller than $N$.

To show (\ref{dinfty}),
let us write the action of the Hecke operators on $Z$ as
\be \label{TLZ}
T'_L Z(q,\bar q) = q^{-c_L L/24}\bar q^{-c_R L/24}
\left(1+ \sum_{m,\bar m>0} \tilde d_{T_L}(m,\bar m) q^m\bar q^{\bar m}\right)\ .
\ee
From the definition
of the Hecke operator it follows that the sum in (\ref{TLZ}) runs
over $m,\bar m \geq  m_0 L$ for some
positive $m_0$.

Introducing $\tilde p =pq^{-\frac{c_L}{24}}\bar q^{-\frac{c_R}{24}}$,
we can write (\ref{symHecke}) as 
\begin{multline}
\exp \left( \sum_{L>0} \frac{1}{L}\tilde p^L + 
\sum_{L>0} \frac{1}{L} \tilde p^L
\sum_{m,\bar m\geq m_0 L} \tilde d_{T_L}(m,\bar m) q^m\bar q^{\bar m} \right) \\
= \left(\sum_{K\geq 0} \tilde p^K \right) \exp \left( 
\sum_{L>0} \frac{1}{L} \tilde p^L
\sum_{m,\bar m\geq m_0 L} \tilde d_{T_L}(m,\bar m) q^m\bar q^{\bar m} \right)\ .
\end{multline}
The sought-after number of states $\tilde d_N(m,\bar m)$ is given
by the coefficient of the term $\tilde p^N q^m \bar q^{\bar m}$.
Since the exponential on the right contributes only non-negative powers of $\tilde p$,
we can restrict the sum on the left to $0\leq K \leq N$. Pulling out
a factor of $\tilde p^N$,
we need to find the coefficient of $\tilde p^0 q^m \bar q^{\bar m}$ of
\be \label{symOrbCoeff}
\sum_{K= 0}^N \tilde p^{-K} \exp \left( 
\sum_{L>0} \frac{1}{L} \tilde p^L
\sum_{m,\bar m\geq m_0 L} \tilde d_L(m,\bar m) q^m\bar q^{\bar m} \right)\ .
\ee
Let us analyze the contribution for a given value of $K$. We consider a
partition $K = \sum_j L_j$, and then pick values $m_j, \bar m_j$ such that
$m=\sum_j m_j$, $\bar m = \sum_j \bar m_j$. Note that this gives an
upper bound for $K$,
\be
K = \sum_j L_j \leq \frac{1}{m_0}\sum_j m_j = \frac{m}{m_0}\ ,
\ee
and similarly for $\bar m$. It follows that for $m$ or $\bar m$
smaller than $m_0 N$, $\tilde d_N(m,\bar m)$ is independent
of $N$. In particular, this shows that the $N\rightarrow\infty$
is well-defined, and can in fact be obtained from the
coefficient of a finite orbifold with $N$ large
enough.

To obtain (\ref{dinfty}), we need to analyze (\ref{symOrbCoeff})
in more detail. For a start, let us consider only the linear expansion
of the exponential in (\ref{symOrbCoeff}), which gives a total
contribution of
\be
\sum_{L=1}^N \frac{1}{L} \tilde d_L(m,\bar m) \ .
\ee
To estimate the contribution of this sum, we can use the fact that
the $T'_L Z$ is again a modular form. This means that we can 
approximate $\tilde d_L$ by the
Cardy formula to get the asymptotic behavior
\be
\tilde d_L(m,\bar m) \sim \exp\left(2\pi 
\sqrt{\frac{c_L L}{6}(m-\frac{c_L L}{24})}\right)
\exp\left(2\pi\sqrt{\frac{c_R L}{6}(\bar m -\frac{c_R L}{24})}\right)\ .
\ee
(We will discuss the validity of this approximation shortly.)
For simplicity we specialize to $c_L=c_R=c$.
We see that this expression is maximized for $L = \frac{24m\bar m}{c(m+\bar m)}$,
where it contributes 
\be \label{growthd}
\tilde d(m,\bar m) \sim e^{4\pi \sqrt{m\bar m}} \ .
\ee
We should now estimate the contributions of all other terms as well
and check if their total contribution makes (\ref{growthd})
grow even faster. The results of the next section show however 
that there is no Hagedorn transition for 
$\beta$ bigger then $2\pi$. This shows that the leading
behavior of (\ref{growthd}) does not change when taking
into account the other contributions.

Note that we have used the Cardy formula for $m = \frac{c}{12}$, 
which is not in its usual range of validity, which requires 
$m \gg c$. Note however that by
the arguments given above, $T'_L Z$ has a gap which 
is parametrically large in $L$. One can show that in this case
the Cardy formula already holds for $m > \frac{c}{12}$ (see
appendix~\ref{app:Cardy} and also \cite{Hartman:2014oaa}
for a more rigorous demonstration of this).

\subsubsection*{2. The free energy $\tilde F(\beta)$}
Let us now discuss in which sense the free energy $\tilde F_N$
is universal in the large $N$ limit. In a first step,
consider the case $\mu=0$.
We need to determine the coefficient of $\tilde p^0$
of (\ref{symOrbCoeff}). 
We can evaluate the 
sum over $m$ by using a saddle point approximation,
\be
\int dm \left(\frac{c_L L}{96 (m-\frac{c_L L}{24})^3}\right)^{1/4} 
e^{2\pi \sqrt{\frac{c_L}{6} L(m-\frac{c_L}{24})}} e^{-\beta m}
= 1\cdot e^{(\frac{4\pi^2}{\beta}-\beta)\frac{c_L L}{24}} + \ldots \ .
\ee
Here we have used the Cardy formula to express $\tilde d(m,\bar m)$,
which is permissible 
for large $L$ since the saddle point is at 
$m_0 = (4\pi^2\beta^{-2}+1)cL/24 + O(1)$. 
The total subleading corrections to all these
manipulations are of order $O(e^{-m_0L\beta})$;
this includes the contribution of terms with
small $L$, where the use of the Cardy formula
is not permitted.

Including the right movers but neglecting all
subleading terms for the moment (\ref{symOrbCoeff})
gives
\be\sum_{K= 0}^N \tilde p^{-K} \exp \left(- \log(1-\tilde p 
e^{(\frac{4\pi^2}{\beta}-\beta)(c_L+c_R)/24})
 \right)\ ,
\ee
so that we can read off from the $\tilde p^0$ coefficient
\be \label{ZNfinite1}
\tilde Z_N(\beta)= \sum_{L=0}^N 
e^{L(\frac{4\pi^2}{\beta}-\beta)(c_L+c_R)/24} = 
\frac{e^{(N+1)(\frac{4\pi^2}{\beta}-\beta)(c_L+c_R)/24}-1}
{e^{(\frac{4\pi^2}{\beta}-\beta)(c_L+c_R)/24}-1}\ . 
\ee
To include the subleading corrections, we use 
the fact that they are $O(e^{-n_0L\beta})$,
so that the result is changed only by multiplication 
with a regular function $e^{-\beta\tilde F^{reg}_N(\beta)}$. In total we thus obtain
\be \label{ZNfinite2}
\tilde Z_N(\beta)= 
\frac{e^{(N+1)(\frac{4\pi^2}{\beta}-\beta)(c_L+c_R)/24}-1}
{e^{(\frac{4\pi^2}{\beta}-\beta)(c_L+c_R)/24}-1}\cdot e^{-\beta\tilde F^{reg}_N(\beta)}\ . 
\ee
The function $\tilde F^{reg}_N(\beta)$ depends on $N$, but it is
bounded uniformly in $\beta$ as long as $\beta$
is not too big. 

From (\ref{ZNfinite2}) we can read off the behavior 
of the free energy in the large $N$ limit. For $\beta<2\pi$
we find that
\be\label{Fhightempphase}
\tilde F_N(\beta) = -N\frac{c_L+c_R}{24}(\frac{4\pi^2}{\beta^2}-1) + O(1)\ .
\ee
On the other hand, for $\beta>2\pi$ we find that
\be \label{Flowtempphase}
\tilde F_\infty(\beta) = \tilde F_\infty^{reg}(\beta)
 + \beta^{-1}\log (1- \exp\left[
\frac{(c_L+c_R)}{24}\left(\frac{4\pi^2}{\beta} - \beta \right)\right])\ ,
\ee
where $\tilde F_\infty^{reg}$ is a regular function whose
specifics depend on the seed theory.

\subsubsection*{3. The general phase diagram} \label{ss:fulldiag}
The above analysis shows that for $N$ going to infinity 
there is indeed a Hagedorn transition at
the temperature $T = \frac{1}{2\pi}$. It also shows 
that for large but finite $N$ the
free energy $\tilde F_N$ diverges as $~N$ above
that temperature. It follows that the rescaled
free energy $f = \lim_{N\rightarrow\infty} N^{-1} F_N$
should be well defined at all temperatures, and that
it should have a first
order phase transition at $T =\frac{1}{2\pi}$.
Its behavior as a function of $T$ 
looks exactly like figure~\ref{fig:freeEnergy}.

We now turn to the case of non-vanishing spin
potential $\mu$. In particular, we want to show
that the full phase diagram of $f(\tau)$ is given
by figure~\ref{fundregion}. To do that,
we use the fact that $Z_N(\tau)$ is modular invariant.
In particular this means that if we know $f(\tau)$
in the fundamental region $\CF_0$, we can obtain
its value for any value of $\tau$ by modular
transformations.

More precisely, we will show that in the $N\rightarrow\infty$
limit, $f(\tau)$ in the fundamental
region $\CF_0$ is given by the contribution of
the vacuum only, \ie
\be \label{fF0}
f(\tau) = -\frac{(c_L+c_R)}{24} \ : \qquad \tau \in \CF_0\ .
\ee
If (\ref{fF0}), it then follows immediately that
the phase diagram is indeed given by figure~\ref{fundregion},
as the free energy is not smoothed out around the
phase transition lines. To show (\ref{fF0}), we
need to show any corrections coming from higher
lying states are negligible. This is equivalent
to showing that $\tilde F_\infty(\tau)$ is
finite in the fundamental region $\CF_0$.

To do this we first need to extract
the partition function $\tilde Z_\infty(\tau,\bar \tau)$.
To do this, we use a trick from \cite{deBoer:1998us}. 
If we define $\tilde p := p q^{-c_L/24} \bar q^{-c_R/24}$, then the $\tilde Z_N$
are the coefficients of the expansion of $\CZ$ in terms of $\tilde p$.
On the other hand we can rewrite $\CZ$ as
\begin{multline}
\CZ = (1-\tilde p)^{-1} \prod_{n>0,m,\bar m\in I}' 
(1-\tilde p^n q^{m/n+nc_L/24} \bar q^{\bar m/n + nc_R/24})^{-d(m,\bar m)\delta^{(n)}_{m-\bar m}}\\
= (1-\tilde p)^{-1} R(\tilde p) \ ,
\end{multline}
where the primed product
omits the factor with $(n=1,m=-c_L/24,\bar m=-c_R/24)$.
Note that we have assumed that the vacuum is unique, \ie $d(-c_L/24,-c_R/24)=1$. If 
the expansion of $R$ is given by
$R(\tilde p) = \sum_k a_k \tilde p^k$, then 
$\tilde Z_N = \sum_{k=0}^N a_k$. It follows that
\be
\tilde Z_\infty = \sum_{k=0}^\infty a_k = R(1) \ ,
\ee
so that the shifted free energy is given by
$\tilde F_\infty$,
\be
-\beta \tilde F_\infty = \log \tilde Z_\infty = 
- \sum_{n>0} \sum_{m,\bar m}' d(m,\bar m)\delta^{(n)}_{m-\bar m} 
\log (1-q^{m/n+nc_L/24}  \bar q^{\bar m/n + nc_R/24}) \\
\ee
It is shown in appendix~\ref{app:convergence} that this expression
indeed converges if $\tau$ is in the standard fundamental region $\CF_0$.
We can then translate this into a statement about the behavior of $f(\tau)$,
from which (\ref{fF0})
and the phase diagram in figure~\ref{fundregion} immediately follow.

\subsection{$SU(N)$ gauge theories: comparison to $d>2$}
Let us briefly explain how our results fit in with the
analysis of $SU(N)$ gauge theories in more than 2 dimensions. In particular
we will compare our findings with the results of
\cite{Sundborg:1999ue,Aharony:2003sx}. It turns out
that although our methods are completely different,
the results are very similar.

In our case the boundary CFT
lives on $T^2= S^1\times S^1$, where the time direction has been
(thermally) compactified. In \cite{Aharony:2003sx} $SU(N)$ CFTs on 
$S^1\times S^{d-1}$ are considered. The rank of the gauge group
then corresponds to $N$ of the orbifold, and the symmetric orbifold
theory corresponds to the free $SU(N)$ gauge theory.

Let us first consider a non-conformal theory which is
asymptotically free, such as pure Yang-Mills.
In the free theory \cite{Aharony:2003sx} also find two phases, 
separated by the Hagedorn temperature $T_H$. In the lower phase,
which is interpreted as the confining phase, the free energy
grows as $N^0$, since the degrees of freedom are (colorless)
glueballs. This agrees perfectly with (\ref{Flowtempphase}).

For higher temperatures, in the deconfined
phase the gluons are the physical degrees of freedom, 
so that the free energy scales as $N^2$. This is 
slightly different from (\ref{Fhightempphase}), where the free energy
grows as $N^1$. We can see where this linear dependence comes
from by considering \eg a system of $Q_1$ D1 and $Q_5$ D5
branes, which in the UV is given by a $U(Q_1)\times U(Q_5)$
gauge theory. When flowing to the IR, the gauge groups 
are eliminated, leaving a symmetric orbifold with $N=Q_1 Q_5$.
Moreover, a careful counting analysis
shows \cite{Hassan:1997ai} that the (1,1) strings and
the (5,5) strings are eliminated by D-term constraints and
gauge fixing, so that only the massless (1,5) modes survive.
The number of degrees of freedom of the resulting effective
CFT is thus proportional to $Q_1Q_5$.

Moreover, according to \cite{Aharony:2003sx} 
in the high-temperature phase,
the free energy grows as $F = N^2 f(T)$ with $F(T) \sim T^d$,
whereas we find $F \sim N T^2$.
In the low-temperature
phase, the density of states should grow Hagedorn-like, 
\ie $d(n) \sim e^{n/T_H}$, which agrees with 
(\ref{growthd}).

Note that our situation is slightly different, as we investigate
a conformal theory. It is clear that on flat space
it cannot have a phase transition, as there is no
energy scale. The energy scale is given by
the radius of the $S^1$ on which we compactify.
In \cite{Witten:1998zw} there is an analogous discussion
of the $d=4, \CN =4$ SYM case, which is also conformal.
Compactified on a finite volume, for purely kinematic reasons one
finds the same behavior as above, which indicates that
(in the large $N$ limit) there are also two phases,
although the low temperature phase has nothing to do
with confinement. In fact, uncompactifying the
$S^3$ shows that $\CN=4$ SYM is always in the unconfined
phase on flat space.

\subsection{Elliptic genus}\label{ss:ellipticGenus}
Our analysis so far was for general bosonic theories
whose partition functions were non-holomorphic.
The original analysis \cite{Dijkgraaf:2000fq} however
used methods that are only applicable for 
holomorphic quantities. One way to obtain
a holomorphic object is to take
a theory with $N=2$ supersymmetry,
and analyze its elliptic genus.
As we will show, the elliptic genus (or rather
the NS partition function obtained from it) has
a phase diagram which is almost the same
as we obtained in the bosonic case, the only
difference being the underlying modular group.

Consider a seed theory with $N=2$ supersymmetry.
Its elliptic genus is defined as 
\be
\chi(q,y)= \Tr_{RR} q^{L_0-c/24}\bar q^{\bar L_0-c/24}y^{J_0}(-1)^F(-1)^{\bar F}
= \sum_{m\geq 0, \ell} c(m,\ell) q^m y^\ell\ .
\ee
Because of the insertion of $(-1)^{\bar F}$, the right
movers only give a constant contribution, the Witten index,
so that the elliptic genus is holomorphic and independent of $\bar \tau$.
For simplicity let us assume that the central charge of the seed
theory is a multiple of six, $c=6k$. The elliptic genus is
then a weak Jacobi form of weight 0 and index $k$.
In particular, it transforms under $SL(2,\IZ)$ as
\be
\chi(\frac{a\tau+b}{c\tau+d},\frac{z}{c\tau+d})
= e^{\frac{2\pi i k c z^2}{c\tau+d}} \chi_1(\tau,z)\ .
\ee
(See \cite{EichlerZagier} for an introduction
to weak Jacobi forms.)
As before, we can write down the generating function for the 
elliptic genus of the symmetric orbifold,
\be
\CZ = \prod_{n>0,m\geq 0, \ell} \frac{1}{(1-p^nq^m y^\ell)^{c(mn,\ell)}}
= \exp\left(\sum_{L>0} \frac{p^L}{L}T'_L\chi(q,y)\right)
\ee
where in this case the Hecke operator is defined as
\be
T'_L\chi(\tau,z)= \sum_{ad=N}\sum_{b=0}^{d-1}\chi(\frac{a\tau+b}{d},az)
=\sum_{ad=N} d \sum_{m\geq0,\ell}c(md,\ell) q^{am}y^{a\ell}
\ee
We now want to specialize the elliptic genus to a partition
function of the theory by specializing the value
of $y$. Since $y$ is the variable that encodes information on the $U(1)$ charge
of the states, we can use its specialization 
to spectrally flow the theory.
In total there are 4 possible spin structures for the left  movers on the torus,
corresponding to RR versus NS sector with and without
the insertion of $(-1)^F$ respectively. The spin
structure which is anti-periodic
in both space and time direction is simply a number,
namely the Witten
index of the theory. The other three spin 
structures are 
functions of $\tau$, and they are interchanged in the 
usual way by acting with $SL(2,\IZ)$.

From the elliptic genus we can then obtain the (left-moving)
twisted partition function in the NS sector,
\be
Z^{\tilde NS,R}(\tau)= \Tr_{NS} (-1)^F q^{L_0-k/4}\ ,
\ee
by half a unit of spectral flow as
\be
Z_N^{\tilde NS,R}(\tau)=(-1)^{kN} q^{kN/4}\chi_N(\tau,z=\tau/2)\ .
\ee
Note that this is strictly speaking not the NS partition
function, but its multiple by the number of right-moving
RR ground states.
Alternatively, we can also obtain the untwisted
NS partition function
\be
Z_N^{NS,R}(\tau) = (-1)^{kN} q^{kN/4}\chi_N(\tau,z=\tau/2+1/2)\ .
\ee
As is expected of a $NS$ partition function, 
$Z_N^{NS,R}$ has simple transformation properties
under the group $\Gamma_\theta=\langle S,T^2 \rangle$ (see \eg
\cite{Gaberdiel:2008xb}). It is thus natural
to expect that its phase diagram is determined
no longer by the fundamental regions of $\Gamma$,
but of $\Gamma_\theta$, such as 
\be \label{fregiontheta}
 \CF_\theta \ : \qquad \tau \in \mathbb{H}\ , \quad |\tau|>1\ ,\quad |\Re(\tau)|<1\ .
\ee
Let us first get a rough picture of the phase diagram from 
general properties of NS partition functions.
Since $Z^{NS,R}(\tau+1)=Z^{\tilde NS,R}(\tau)$,
the phase diagrams of the two functions
are simply shifted.
The high temperature behavior
of $Z_N^{NS,R}$ is obtained as in the bosonic case,
\be
Z_N^{NS,R}(\tau)= e^{2\pi i kN/4\tau}+\ldots\ ,
\ee
which shows that there is again a phase transition
at $T=\frac{1}{2\pi}$. This is not surprising, as this
partition function simply counts the bosonic and fermionic
states in the NS sector. The behavior
of the twisted partition function is more interesting
because of the insertion of $(-1)^F$, which 
can lead to cancellations between fermionic and
bosonic states. In fact, 
since 
\be
Z^{\tilde NS,R}(-1/\tau) = Z^{R,R}(\tau)\ ,
\ee
there is no divergence in the high
temperature case, as 
\be
Z^{\tilde NS,R}(0)=Z^{R,R}(i\infty)= N_{RR}\ ,
\ee
the number of Ramond-Ramond ground states. 
For a symmetric orbifold theory with finite $N$ the
$Z^{\tilde NS,R}$ thus does not diverge as
we go to arbitrarily high temperatures.

To get a rough idea of what will happen in 
the $N\rightarrow\infty$ case, let us find a lower
bound for the number of these states by evaluating the
elliptic genus for $z=0$. This gives
\be
\CZ = \left(\prod_{n>0}(1-p^n)^{-\sum_\ell c(0,\ell)}\right)
\ee
whose asymptotic behavior is
\be\label{FRR}
Z_N^{RR} \sim e^{2\pi \sqrt{\sum_{\ell} c(0,\ell) N/24}}\ .
\ee
Assuming that total number of Ramond-Ramond ground states
has the same behavior as this lower bound, we see that
asymptotic behavior for $N\rightarrow\infty$ is quite different.
In particular this indicates 
that the rescaled free energy $f$ will not have 
a phase transition along the imaginary axis, which 
agrees with the phase diagram outlined above. The exact analysis
below will confirm this.

Let us now analyze the free energy of the twisted
NS partition function $\tilde Z^{\tilde NS,R}$ in the strict $N=\infty$ case
in detail.
As in the bosonic case, we introduce the shifted variable $\tilde p=pq^{-k/2}$
and then factor out the lowest lying term $(1-p)^{n+1}$. In general
however there will be more than single ground state, so that
$n>0$. We would however still like to obtain the result by
setting $p=1$ in the remainder of the expression.
As a more careful analysis shows
(see appendix~\ref{app:ellipticgenus}),
the fact that there is more than one ground state
adds a term $n \log N$ to the free energy. Even though
this terms makes $\tilde F_\infty$ diverge,
it is a subleading contribution to the rescaled free
energy $f$. To obtain the phase diagram, we can
thus neglect it and only consider
\be \label{FNSRtw}
\log \tilde Z_\infty^{\tilde NS,R}(\tau)
= \sum_{a,n}\frac{1}{a}q^{ank/2}\sum_{m,\ell}' c(mn,\ell)q^{a(m+\ell/2)}
= \sum_{a,n}\frac{1}{a}q^{ank/2}\sum_{b=0}^{n-1} \chi(\frac{a\tau+b}{n},a\frac{\tau}{2})
\ee
for the twisted partition function.

We want to show that (\ref{FNSRtw}) converges in some region of the upper
half plane. Since $Z^{\tilde NS,R}$ is invariant under
the modular subgroup $\Gamma_\theta = \langle S, T^2\rangle$,
we can assume that $\Re(\tau_1)\leq 1$. The problem is then essentially
the same as in the bosonic case: we need to show that 
the Hecke transform of $\chi$ appearing in (\ref{FNSRtw})
does not grow too quickly, so that it can be suppressed 
by the prefactor $q^{akn/2}$.

Since we are interested in the behavior for $n$ large,
it is natural again to use an $S$ transformation
to obtain
\be
\chi(\frac{a\tau+b}{n},a\frac{\tau}{2})= \chi(-\frac{n}{a\tau+b},\frac{na\tau/2}{a\tau+b})
e^{-2\pi i k\frac{n(a\tau/2)^2}{a\tau+b}}\ .
\ee
In the bosonic case the leading contribution was then simply
the vacuum. For the elliptic genus, we need to find
which term $q^m y^\ell$ gives the dominant contribution.
Not surprisingly, it turns out that we only need to worry
about $a=1$, and the dominant term has
$m=0$ and $\ell= \pm k$, the highest allowed value of $\ell$,
and $b=\pm 1)$ --- see appendix~\ref{app:ellipticgenus}.
The total contribution in (\ref{FNSRtw}) is
\be
\exp\left[2\pi \tau_2 kn\left(-\frac{1}{4}+\frac{1}{4|\tau\pm1|^2}\right)\right]\ ,
\ee
so that (\ref{FNSRtw}) converges if and only if $|\tau+1|>1$ and
$|\tau-1|>1$. The the free energy of the untwisted partition 
function $Z^{NS,R}$ thus converges
in the fundamental region (\ref{fregiontheta}), and $f(\tau)$
in this region is given by $-\frac{k}{2}$.


\subsection{Order parameters}
We would now like to analyze the contribution of the different
twist sectors to the free energy. We expect that at low temperature,
the main contribution will come from the untwisted sector
which contains the vacuum. The higher the temperature, 
the more important the contributions of the twisted sectors
will be. To make these statements more precise, 
we want to measure the contributions by
constructing a family of order parameters.
Not surprisingly, we will find that these order parameters
undergo a phase transition at the Hagedorn temperature.\footnote{Strictly speaking, the canonical ensemble is 
no longer well-defined above this temperature. It is still
possible however to work with finite $N$ and then
take the limit $N\rightarrow\infty$ 
of the expectation values of the order parameters.}
Their critical exponent turns out to be universal.

We want to define order parameters measuring the contribution
of different twist sectors whose expectation
values that can be easily evaluated.
As was mentioned before, a twist sector in the theory is given
 by the conjugacy classes, so that
\be
(1)^{N_1} (2)^{N_2} \cdots (s)^{N_s}\ , \qquad \sum_n nN_n =N\ .
\ee
Before projection, a particular sector is therefore the tensor product
of cycles of length $n$. It is therefore natural to define
observables whose action factorizes over those cycles.
We define a family of order parameters $P_g$ who act on
a cycle of length $n$ by multiplication by $g(n)$,
where $g$ is an arbitrary function of $n$.
They then act on states of the twisted sector as
\be \label{gorder}
P_g |\{N_n\}\rangle = \prod_n g(n)^{N_n} |\{N_n\}\rangle\ .
\ee
The advantage of having an observable of the form
(\ref{gorder}) is that it is straightforward
to obtain the generating function of its expectation
values by repeating the original argument.
Define $Z_g(S^N \CC_1; q, \bar q)$
as the trace of the symmetric orbifold theory with $P_g$ inserted.
Since by construction $P_g$ factorizes over all twisted sectors,
we can repeat the original derivation to obtain the generating function,
\begin{eqnarray}
\CZ_g &=& \sum_{N\geq 0} p^N Z_g(S^N \CC_1;q, \bar q)\nonumber \\
&=& \sum_{N\geq 0} \sum_{\sum nN_n=N} \prod_{n>0} p^{nN_n} g(n)^{N_n} 
Z(S^{N_n} H^{Z_n}_{(n)}; q,\bar q) \nonumber \\
&=& \prod_{n>0} \sum_{K\geq 0} (g(n)p^n)^K Z(S^K H^{Z_n}_{(n)}; q,\bar q) 
\end{eqnarray}
where we have used the multiplication property of $P_g$ in the second line. 
We can then use the same reasoning as in the original argument to
sum over $K$ to get
\be
\CZ_g = \prod_{n>0} \prod_{m,\bar m} (1-g(n)p^n q^m \bar q^{\bar m})^{-c_n(m,\bar m)}\ .
\ee
By using the same trick as in section~\ref{ss:fulldiag} we can
then extract the expectation values of these order parameters
for $N =\infty$, 
\be
\langle P_g \rangle_{\infty} = \frac{\tilde Z^g_\infty}{\tilde Z_\infty}\ .
\ee

Let us now evaluate the expectation value for some of those
order parameters.
The contribution of the untwisted sector can be extracted by picking
$g(1)=1,\ g(n>1)=0$:  
\be \label{Putw}
\langle P_{\textrm{utw}} \rangle_{\infty}
= \tilde Z_\infty^{-1} \prod_{(m,\bar m)>(-c_L/24,-c_R/24)} 
(1-q^{m+c_L/24} \bar q^{\bar m + c_R/24})^{-d(m,\bar m)} \ .
\ee
We thus see that the contribution of the untwisted sector starts as 1 for $T=0$, decreases
with growing temperature and vanishes once we cross the Hagedorn transition
point. In particular this shows that $\langle P_{\textrm{utw}} \rangle_{\infty}$ cannot
be an analytic function of $\beta$.
More precisely,
the infinite product in (\ref{Putw}) is regular, so 
that the behavior of $\langle P_{\textrm{utw}} \rangle_{\infty}$
near the critical temperature is determined by
the behavior of $\tilde Z_\infty^{-1}$. Using (\ref{Flowtempphase})
we see that
\be \label{critexp}
\langle P_{\textrm{utw}} \rangle_{\infty}(T)
\sim (T_H-T)^{1}\ , 
\ee
\ie the critical exponent of the order parameter is 1. Note that
although the vanishing of $Z^{-1}$ above the critical 
temperature is a generic feature of Hagedorn transitions,
the exact value of the critical exponent depends on the theory. 
The value given in (\ref{critexp}) is thus a statement
on the subleading behavior of the growth of the number
of states in symmetric orbifold theories.

\begin{figure}
\begin{center}
\includegraphics[width=0.45\textwidth]{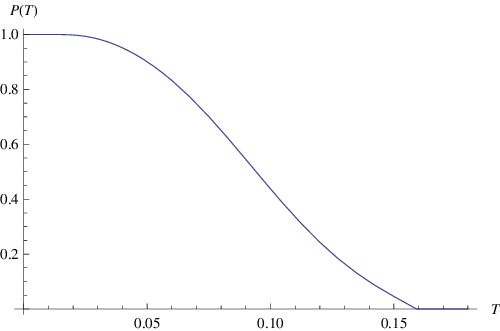}
\includegraphics[width=0.45\textwidth]{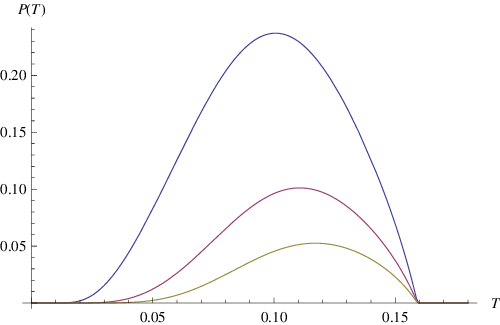}
\end{center}
\caption{\label{orderParameter} 
The free boson at self-dual radius:
$\langle P_{\textrm{utw}}\rangle_\infty$ as a function of
$T$ (left), $\langle P_{(n)}\rangle_\infty$ for $n=2,3,4$ (right).}
\end{figure}

To estimate the contribution of cycles of length $n$, we choose
$g(1)=g(n)=1,\ g=0$ else. We include contributions of
cycles of length 1 to ensure that for any finite $N\geq n$ there
is a contribution. Subtracting the contribution of the
untwisted sector we get
\begin{multline}
\langle P_{(n)}\rangle_\infty=\langle P_{(1),(n)}\rangle_\infty - 
\langle P_{(1)}\rangle_\infty
= 
\prod_{n'>1,n'\neq n,m,\bar m} 
(1-q^{m/n'+n'c_L/24} \bar q^{\bar m/n' + n'c_R/24})^{d(m,\bar m)\delta^{(n')}_{m-\bar m}}\\
- \prod_{n'>1,m,\bar m} 
(1-q^{m/n+n'c_L/24} \bar q^{\bar m/n' + n'c_R/24})^{d(m,\bar m)\delta^{(n')}_{m-\bar m}}
\end{multline}
We know that this vanishes for $T=0$ and $T=\frac{1}{2\pi}$, 
so that there must be a 
maximum somewhere in between. Figure~\ref{orderParameter}
shows $\langle P_{(n)}\rangle_\infty$ for a free boson at 
the self-dual radius. As expected, the higher 
twisted sectors achieve their maximal contribution
at a higher temperature.


\section{General families}
\setcounter{equation}{0}
\subsection{Moving away from the orbifold point}
Our analysis so far was perfectly appropriate for 
problems in CFT. If we want to apply it to holographic
duals of gravity theories, we need to modify it somewhat. 
In general, the holographic dual
is not expected to be at the symmetric orbifold
point, but rather at some other point in the
moduli space of the theory where
the orbifold has been resolved \cite{Seiberg:1999xz}. 
In the D1-D5 system for instance this means
one has to deform the theory by exactly marginal
operators in the twist 2 sector \cite{David:2002wn}.
Even though there has been recent progress
in performing computations in such deformed theories
\cite{Pakman:2009zz,Avery:2010er,Avery:2010hs},
to compute the spectrum of the deformed theory
is still a very difficult task.

Let us thus turn the problem around. Instead 
of trying to deform the symmetric orbifold, we will 
analyze the phase diagram of general families of CFTs
to see if they can possibly serve as a holographic dual
to gravity theories\footnote{See \cite{ElShowk:2011ag} for
a recent discussion of this question using
different methods.}. 
The most restrictive point of view one can take is
that the phase diagram of such a family $\{\CC^{(N)}\}_{N\in\mathbb{N}}$
should look exactly like figure~\ref{fundregion},
and that their free energy should be given
by figure~\ref{fig:freeEnergy}.
For this to happen the free energy has to satisfy
\be
\lim_{N\rightarrow\infty} \frac{1}{N} \tilde F_N(\beta) =0\ , \label{criterion1}
\ee
for all $\beta > 2\pi$.
To rewrite this in terms of the states of the theory,
use
\be
Z_N(\beta) = e^{-\beta\frac{(c_L+c_R) N}{24}}\left( 1+ \sum_{m,\bar m} 
\tilde d_N(m,\bar m) e^{-\beta(m+\bar m)}\right) \ ,
\ee 
so that (\ref{criterion1}) becomes
\be \label{criterion3}
\lim_{N\rightarrow\infty} \frac{1}{N}\log \left( 1+ \sum_{m,\bar m} 
\tilde d_N(m,\bar m) e^{-\beta(m+\bar m)}\right) = 0 .
\ee
Physically,
the interpretation of (\ref{criterion3}) is clear. If the
free energy is to look like figure~\ref{fig:freeEnergy}, 
then the low energy behavior must be given
by the vacuum contribution alone. Although the
contribution of the states above the vacuum is exponentially 
suppressed, there may be a lot of them, which is potentially
enough to produce a Hagedorn behavior even at low temperature.
This would then give a non-vanishing contribution to
the free energy $f$.
In particular it is possible that they can 
smooth out the Hawking-Page transition.
Of course a gravity theory might also have a more complicated 
behavior than figures~\ref{fig:freeEnergy} and \ref{fundregion}.
Given such a behavior, one could again find criteria
that holographic duals have to satisfy. For simplicity
however let us restrict to (\ref{criterion3}).

Unfortunately (\ref{criterion3}) is not very transparent. 
In particular, some of the coefficients $\tilde d_N(m,\bar m)$ 
we can choose freely, whereas others are fixed by
modularity. To make (\ref{criterion3}) more transparent
we return to the example of holomorphic theories.

\subsection{Holomorphic partition functions} 

For holomorphic partition functions one can render 
(\ref{criterion1}) more transparent. Holomorphic partition functions appear
in two contexts: for holomorphically factorisable theories and for
chiral theories, where there are no right-moving fields and thus
the holomorphic part by itself is the full partition function.

Let us thus consider a chiral theory of central charge $c=24N$. Its
partition function is a modular function with a pole of order $N$
at $q=0$. Such a function is uniquely determined by its polar and
constant part, so that we need to fix only a finite number
of coefficients $d(n)$. One way to see this is to use the Rademacher expansion
of such a function. 

For our purposes it will be more useful
to follow \cite{Witten:2007kt}. 
Let 
\be
J(q) = q^{-1} + 196884q + \dots
\ee
be the (modular invariant) partition function of the Monster CFT.
It is related to the usual $j$-invariant by $J(\tau)=j(\tau)-744$.
Acting with the Hecke operator $T'_n$ we get
\be
T'_n J(q) = q^{-n} + O(q) \ .
\ee
It thus follows that we can write any holomorphic partition function as
\be
Z_N(\tau) = \sum_{n=0}^N \tilde d_N(n)T'_{N-n} J(\tau) \ ,
\ee
\ie we can express all $\tilde d_N(n)$ in terms of the first $N$ coefficients.

One particular family of CFTs of interest are the proposed extremal CFTs of 
\cite{Witten:2007kt}. Their central charge is given by a multiple of 24,
and their partition function factorizes into a holomorphic and antiholomorphic
part. They are called extremal because they contain the minimal operator
content compatible with modularity.
In this case the $\tilde d_N(n)$ are given by
the Virasoro descendants of the vacuum, 
\be
\tilde d_N(n) = \left[ \prod_{l=2}^\infty \frac{1}{1-q^l} \right]_{q^n} \ , 
\quad n = 0,\ldots, N \ . \label{descendants}
\ee
The fact that we can write down modular invariant partition
functions is of course not enough to guarantee the existence
of extremal CFTs. It is in fact still an open question whether
they exist for central charge bigger than 24 
\cite{Gaberdiel:2007ve,Gaberdiel:2008pr,Gaiotto:2007xh,Gaberdiel:2010jf}.
We will simply use their partition functions as a template
for our analysis. In fact it will turn out that one can 
greatly relax extremality condition on the spectrum and
still get Hawking-Page transitions.

The proposed extremal CFTs are supposed to be dual to pure gravity on $AdS_3$,
so that extremal partition function should 
exhibit Hawking-Page phase transitions. This was indeed shown in 
\cite{Maloney:2007ud} by showing that for $N\rightarrow\infty$ there
is a Lee-Yang type condensation of zeros of the partition function
on the circle $|\tau|=1$. In view of (\ref{criterion3}) this is no
surprise, as extremal partition functions contain the fewest states possible.

One can relax the extremality condition quite a lot
and still get Hawking-Page transitions.
More precisely, 
assume that for $n\leq N$ the coefficients grow as
\be
\tilde d(n)= e^{2\pi \alpha n} e^{o(n)} \ ,
\ee
with $o(n)$ some function such that $\lim_{n\rightarrow\infty} n^{-1} o(n)$.
For simplicity we will also assume that $o(n)$ is monotonically growing.
This function $o(n)$ may encode polynomial prefactors, but also
Virasoro descendants of 
primaries, since (\ref{descendants}) grows only like $\exp (\pi \sqrt{2n/3})$.

We can easily find a lower bound for the free energy by considering
the contribution of states at level $N$, obtaining 
\be \label{lowboundF}
\log \tilde F(\beta) \geq (2\pi\alpha-\beta) N + o(N)\ .
\ee
It follows directly that for $\alpha>1$ (\ref{criterion3}) is violated 
so that there is no transition. 
On the other
hand we will show that for $\alpha \leq 1$ there is a Hawking-Page
transition.

To do this first note that for $n$ prime
\be
T'_n J(i) = J(in) + \sum_{b=0}^{n-1} J((i+b)/n)
\leq J(in) + \sum_{b=0}^{n-1} J(i/n) = (n+1)J(in)\ ,
\ee
where the inequality comes from the fact that the coefficients of $J(q)$ 
are real and positive. We can bound this expression by using
\be
J(in) \leq e^{2\pi n} + J(i)\ , \label{bound1}
\ee
which again follows from the positivity of the coefficients. If $n$ is not
prime, then we see from (\ref{Hecke}) that 
there are at most $n$ such sums with $d$ terms each. Since
$d\leq n$ we can again bound their contribution by (\ref{bound1}), getting
in total
\be
T'_n J(i) \leq n^2 (e^{2\pi n} + J(i))\ .
\ee
For $\alpha \leq 1$ we can thus estimate
\be
e^{-2\pi N} Z_N(i) \leq e^{-2\pi N} 
\sum_{n=0}^N \tilde d(n) (N-n)^2( e^{2\pi (N-n)} + J(i)) \leq 
N^3 e^{o(N)} (J(i)+1) \ ,
\ee
which satisfies 
(\ref{criterion1}). Note that this is compatible with the analysis
of \cite{Maloney:2007ud}, where a condensation of zeros was 
found for the weaker bound $\alpha\leq 0.61$ and $o(n) = 0$.

In principle one could also perform the same type of analysis
for the elliptic genus, \ie classify $N=2$ SCFT which
exhibit Hawking-Page transitions. Note however that
unlike the holomorphic case, not every combination
of polar terms can be completed to a weak Jacobi form
(see \cite{Gaberdiel:2008xb} in the context of
pure supergravity). The coefficients of polar terms
will thus have to satisfy certain constraints.


\subsection*{Acknowledgements}
It is a pleasure to thank Costas Bachas and Jan Troost for initial collaboration
on this project.
I would also like to thank Miranda Cheng, Terry Gannon, Amit Giveon, 
Sean Hartnoll, Alex Maloney, 
Shiraz Minwalla, Shlomo Razamat and Andy Strominger for helpful discussions,
and Alex Maloney and Matthias Gaberdiel for useful remarks on the draft.
My research is supported by a John A.~McCone Postdoctoral Fellowship. 

\appendix
\renewcommand{\theequation}{\Alph{section}.\arabic{equation}}
\setcounter{equation}{0}

\section{Appendix}
\subsection{A very brief remark on the Cardy formula}\label{app:Cardy}
In this appendix we will briefly rederive the Cardy formula \cite{Cardy:1986ie},
concentrating on its range of validity. 
The main purpose is
to show that if the partition function has a large gap,
then the Cardy formula is valid even for small values of $n$.
For simplicity we restrict to the holomorphic case.
\footnote{See \cite{Loran:2010bd} for a discussion why right and left
movers factorize for large $m$ and $\bar m$.}
Let $Z(q)= \sum_{n} d(n) q^n$ be the modular invariant partition
function. We can obtain $d(n)$ by contour integration,
\be
d(n) = \oint_C dq q^{-n-1} Z(q) \ ,
\ee
where $C$ is an arbitrary closed contour that encloses the origin.
We choose $q=e^{2\pi i \tau}$, $\Im\, \tau = \sqrt{\frac{c}{24 n}}$,
$\Re\, \tau \in [-\pi, \pi]$.
We can then use modularity to write
\be
\int d\tau e^{-2\pi in \tau} Z(-1/\tau)
= \int d\tau e^{-2\pi in \tau}  e^{\frac{c}{24} 2\pi i/\tau}
(1+ O(e^{-in_0 L/\tau}))
\ee
The point is that for large $L$, the second term is negligible
as long as $\tau$ remains of order 1. This is the case if
$n \sim c$. We can then perform the usual saddle point approximation
for the first term and obtain a maximum at 
$\tau_0 = i\sqrt{\frac{c}{24 n}}$. Moreover the second derivative
at $\tau_0$ is negative, real and proportional to $n^{3/2}c^{-1/2}$, so that 
for large $c$ the saddle point approximation is valid. In total we
thus get
\be \label{Cardy}
d(n) \sim \left(\frac{c}{96n^3}\right)^{1/4} e^{2\pi \sqrt{cn/6}}\ .
\ee

\subsection{Convergence of the free energy}\label{app:convergence}
The free energy is given by 
$\tilde F_\infty$,
\begin{multline}
\beta^{-1} \tilde F_\infty = \log \tilde Z_\infty = 
- \sum_{n>0} \sum_{m,\bar m}' d(m,\bar m)\delta^{(n)}_{m-\bar m} 
\log (1-q^{m/n+nc_L/24}  \bar q^{\bar m/n + nc_R/24}) \\
= \sum_{n>0} \sum_{m,\bar m}' d(m,\bar m)\delta^{(n)}_{m-\bar m} 
\sum_{a=1}^\infty a^{-1} q^{a(m/n+nc_L/24)}  \bar q^{a(\bar m/n + nc_R/24)} . \label{logZ}
\end{multline}
We want to show that this expression converges 
if $\tau$ is in the fundamental domain $\CF_0$,
$|\tau| > 1, |\Re(\tau)| < \frac{1}{2}$.

Some brief comments on convergence are in order. For fixed $n$, 
the sum over $m,\bar m, a$ converges absolutely, as can be seen 
by using $-\log (1-x) \leq \frac{x}{1-x}$ and the fact
that the asymptotic behavior of $d(m,\bar m)$ is determined
by the Cardy formula. This allows us
to exchange the summation over $m$ and $a$.
Moreover, the divergent behavior of the sum is determined by 
large values of $n$ only, so that for the rest of the argument 
we can always assume that $n$ is large. For a start we 
will concentrate on the term $a=1$, as we will argue later
on that this is the only important term.

Using
\be
\delta_{m-\bar m}^{(n)} = \frac{1}{n}\sum_{k=0}^{n-1} e^{2\pi i(m-\bar m)k/n}
\ee
we rewrite the $n$-th term of (\ref{logZ}) as
\be \label{termn}
\frac{1}{n}\sum_{k=0}^{n-1} \sum_{m,\bar m} d(m,\bar m) e^{2\pi i(m-\bar m)k/n}
 q^{(m/n+nc_L/24)}  \bar q^{(\bar m/n + nc_R/24)}
 =\frac{1}{n}\sum_{k=0}^{n-1} Z_1(\frac{\tau+k}{n})q^{nc_L/24}q^{nc_R/24}\ .
\ee
We now claim that for $n$ large enough and $\tau \in \CF_0$, 
$|Z_1(\tau_k)| \leq |Z_1(\tau_0)|$ (where $\tau_k=(\tau+k)/n$).
To show this, let us map each $\tau_k$ to its image $\hat \tau_k$
in $\CF_0$. Clearly $\Im(\hat\tau_0)= \Im(\tau)\frac{n}{|\tau|^2}$.
It is then straightforward to show that all the other $\hat \tau_k$
have smaller imaginary part\footnote{To see this note that
by choosing a suitable summation range for $k$, $\tau_k$ is
always in the strip $|\Re(\tau_k)|<\frac{1}{2}$ with $|\Re(\tau_k)|>|\Re(\tau_0)|$.
After the first $S$ transform its imaginary part is thus
smaller than that of $\hat\tau_0$, and further
modular transformations do not change this.}.
Since for $\Im(\tau)$ big enough $Z_1(\tau)$ is monotonically
increasing, this shows the claim.

Using that $\Im(\hat\tau_0)$ is very large, we can thus bound
(\ref{termn}) by
\begin{multline}
\exp(2\pi \frac{c_L \Im(\hat \tau_0) + c_R\Im(\hat \tau_0)}{24})
\exp(-2\pi \frac{n(c_L \Im(\tau)+c_R\Im(\tau)}{24})\\
=\exp(2\pi \frac{n(c_L+c_R)\Im(\tau)}{24}(\frac{1}{|\tau|^2}-1))\ .
\end{multline}
The coefficient of $n$ is negative if
\be
|\tau|^2>1,
\ee
in which case the sum converges. Including the terms with $a>1$ simply
changes the temperature to $a\tau$, and the resulting sums
in $n$ fall off exponentially in $a$, so that the sum
over $a$ converges.

\subsection{Degenerate ground states}
Let us briefly discuss the case of degenerate ground states.
If there are $n+1$ ground states then the generating function
is of the form
\be
F(p)=\sum_{N\geq 0} d_N p^N = \frac{1}{(1-p)^{n+1}}f(p)
\ee
with $f(p) = \sum_{k\geq0} a_k p^k$. Moreover define
$f_N(p)=\sum_{k=0}^N a_k p^k$. We know that $f(p)$
and its first $n$ derivatives converge for $p=1$.
This follows from the fact that the derivative
of $\log f(p)$ converges, which can be shown 
by the same type of analysis used to show
convergence of $f(1)$.
We then have
\be
d_N = \frac{N^n}{n!} (f_N(1) + O(N^{-1}))\qquad N\rightarrow\infty\ . 
\ee
To see this, note that 
\be
d_N = \sum_{k=0}^N a_k \binom{N-k+n}{n}= \frac{1}{n!}\frac{d^n}{dp^n}\left(p^{N+n}f_N(p^{-1})\right)|_{p=1}
= \frac{N^n}{n!}f_N(1)+O(N^{n-1})
\ee
From what we have said above $f_N$ at its derivatives at $p=1$
are bounded.
It thus follows that the degeneracy of the ground states only
contributes a polynomial prefactor. In particular,
its contribution to the free energy is
subleading and can be neglected after rescaling
by $N$.

\subsection{The dominant contribution for the elliptic genus}\label{app:ellipticgenus}
We need to find the term with the biggest contribution in
the expansion of
\be\label{ellgenusexpansion}
\sum_{b=0}^{n-1}
\chi(-\frac{n}{a\tau+b},\frac{na\tau/2}{a\tau+b})
e^{-2\pi i k\frac{n(a\tau/2)^2}{a\tau+b}}\ .
\ee
Use
\ba
\Im \left(-\frac{n}{a\tau+b}\right) &=& \frac{na}{|a\tau+b|^2}\tau_2 \ ,\\
\Im \left(\frac{na\tau/2}{a\tau+b}\right) &=& \frac{ban}{2|a\tau+b|^2}\tau_2\ ,\\
\Im \left(\frac{n(a\tau/2)^2}{a\tau+b}\right) 
&=& \frac{na\tau_2}{4}\left(1-\frac{b^2}{|a\tau+b|^2}\right) \ .
\ea
The exponent of the absolute value of the term corresponding to
 $q^m y^\ell$ in (\ref{ellgenusexpansion}) is 
\be
-2\pi\tau_2  na\left( \frac{m+b\ell/2+\frac{k}{4}b^2}{|a\tau+b|^2} -\frac{1}{4} \right)\ .
\ee
Let us now find the minimum of the fraction in the above expression.
It is clear that $|\ell|$ must attain its maximal value
$\sqrt{4mk+k^2}$, so that 
\be
\frac{4m\pm2b\sqrt{4mk+k^2}+kb^2}{4|a\tau+b|^2}
= k\frac{(b\pm\sqrt{1+4m/k})^2-1}{4|a\tau+b|^2}
\ee
From the form of this expression we see that again 
we can concentrate on the case $a=1$.
Remembering that $|\tau_1|<1$, it is then straightforward to see that 
the exponent is minimal for $m=0$ with $b=\pm1$,
and its contribution is
\be
\exp\left[2\pi \tau_2 kn\left(\frac{1}{4}+\frac{1}{4|\tau\pm1|^2}\right)\right]\ .
\ee


\end{document}